\documentclass[preprint,review,12pt]{elsarticle}

\usepackage{amssymb}

\usepackage{amsmath}
\usepackage{epsfig}
\usepackage{graphicx}
\usepackage[colorinlistoftodos,prependcaption,textsize=tiny]{todonotes}
\usepackage{color,soul}

\begin{document}
	
	\begin{frontmatter}
		\title{Electromagnetic dipole and quadrupole moment interaction of $\Omega^{\pm}$ baryons with bent crystals at the LHC}

		\author{V.G.~Baryshevsky}
	\address{Research Institute for Nuclear Problems, Belarusian State University,
	Bobruiskaya 11, 220030 Minsk, bar@inp.bsu.by, v\_baryshevsky@yahoo.com}
		
		\begin{abstract}
Spin rotation of $\Omega^{\pm}$ baryons channeling in a bent crystal at the LHC in addition to EDM measurement open possibility to study $\Omega^{\pm}$ baryons electric quadrupole moment.
Equations for spin and tensor polarization rotation are presented along with the estimations study for values of the effects under discussion.
		\end{abstract}

		
	\end{frontmatter}

   \section{Introduction}
Recently, the experimental approach was proposed \cite{Phys_1,Phys_2,VG_3,VG_4} to search for the electrical dipole moments (EDM) of charged short-lived heavy baryons and $\tau$-leptons using bent crystal at LHC.
According \cite{VG_x,VG_6,VG_7}, the same approach gives a unique possibility for investigating the P-odd T-even and P-odd T-odd (CP-odd) interactions of short-lived baryons ($\tau$-leptons) with electrons and nuclei.
Possibility to measure the quadrupole moment of $\Omega^\pm$ baryon channeling in a straight crystal 
was also demonstrated  \cite{Phys_5,Phys_6}.
In present paper equations for $\Omega^{\pm}$ spin and quadrupole spin-tensor (second rank tensor) are presented.
It is shown, that spin rotation of $\Omega^{\pm}$ baryons channeling in bent crystal in addition to EDM measurement open possibility to study $\Omega^{\pm}$ baryons electric quadrupole moment at the LHC.

\section{Spin interactions of relativistic $\Omega^{\pm}$ baryon with crystals}

For high-energy particles moving in intracrystalline electromagnetic fields the particle wavelength $\lambda$ is much smaller as compared to the typical field nonuniformity: $\lambda\ll R_{sh}$, where $R_{sh}$ is the shielding radius.
This fact enables applying the quasi-classical approximation. As a result to study particle spin evolution in the electromagnetic fields inside the crystal one can use
the Thomas-Bargmann-Michael-Telegdi (T-BMT) equations. The T-BMT equation describes spin motion in the rest frame of the particle, wherein spin is described by three component vector $\vec{S}$ \cite{Phys_13}. The term ''particle spin'' here means the expected value of the quantum mechanical spin operator $\hat{\vec{S}}$, hereinafter the symbol marked with a ''hat'' means a quantum mechanical operator.

Modification of the T-BMT equation allowing for the possible existence of non-zero electric dipole moment (EDM) for a particle moving in a crystal were studied in  \cite{Phys_1,Phys_2,VG_3,VG_4,VG_x,VG_6,VG_7,Phys_5,Phys_6}.

Futher description will be done for a relativistic particle, which moves in the non-magnetic crystal. Let us suppose particle Lorentz-factor $\gamma$ to be high: $\gamma \gg 1$. In this case spin motion is described by:
\begin{equation}\label{eq1}
\frac{d\vec{S}}{dt}=[\vec{S}\times\vec{\Omega}_{magn}],
\end{equation}
\begin{equation}\label{eq2}
\vec{\Omega}_{magn}=-\frac{e(g-2)}{2mc}[\vec{\beta}\times\vec{E}],
\end{equation}
where $\vec{S}$ is the spin vector, $t$ is the time in the laboratory frame, $m$ is the mass of the particle, $e$ is its charge, $\vec{\beta}=\frac{\vec{v}}{c}$, where $\vec{v}$ denotes the particle velocity, $\vec{E}$ is the electric field at the point of particle location in the laboratory frame, and $g$ is the gyromagnetic ratio. By definition, the particle magnetic moment is $\mu=(eg\hbar /2mc)S$.\\
If a particle possesses an intrinsic electrical dipole moment, then the additional term, describing spin rotation induced by the EDM, should be added to (\ref{eq1}):
\begin{equation}\label{eq3}
\frac{d\vec{S}_{EDM}}{dt}=[\vec{S}\times\vec{\Omega}_{EDM}],
\end{equation}
\begin{equation}\label{eq4}
\vec{\Omega}_{EDM}=\frac{D}{S \hbar}\vec{E}_{\perp}=\frac{ed}{S\hbar}\vec{E}_{\perp},
\end{equation}
where $D=ed$ is the electric dipole moment of the particle.\\
As a result, motion of a particle spin due to the magnetic and electric dipole moments can be described by the following equation:
\begin{equation}\label{eq5}
\frac{d\vec{S}}{dt}=[\vec{S}\times\vec{\Omega}_{magn}]+[\vec{S}\times\vec{\Omega}_{EDM}].
\end{equation}

The above equation is not applicable for studies dedicated for EDM search of $\Omega^\pm$ baryon, because its spin $S=\frac{3}{2}>\frac{1}{2}$ and $\Omega^\pm$ baryon can possess quadrupole moment.
This is the reason to generalize (\ref{eq5}) by adding the term, which describes interaction of relativistic particle quadrupole moment with nonuniform electric field of the crystal \cite{VG_7,Phys_6}.
The corresponding term describing such interaction was obtained in  \cite{Phys_5,VG_12,VG_13} and reads as follows:
\begin{equation}\label{eq6}
\frac{d\hat{S}_i}{dt}= \frac{eQ}{3\hbar}\varepsilon_{ikl}\varphi_{kn}\hat{T}_{ln},
\end{equation}
where $\hat{S}_i$ is the operator of the particle spin projection; repeated indices imply summation:
\begin{equation}\label{eq7}
\hat{T}_{ln}=\frac{3}{2S(2S-1)} \left\{\hat{S}_{ln}-\frac{2}{3}S(S+1)\delta_{ln}\right\},
\end{equation}
where $\hat{T}_{ln}$ is the second rank tensor \cite{Varshalovich}, Q is the particle quadrupole moment, 
$\hat{S}_{ln}=\hat{S}_i\hat{S}_n+\hat{S}_n\hat{S}_i$, 
$\varphi_{kn}=\frac{\delta^2\varphi}{\delta x_k\delta x_n}$, 
the electric potential $\varphi$ is evaluated in the point of particle location in the crystal at the instant $t$,
$\varepsilon_{inl}$ is the totally antisymmetric unit tensor.

Let us add (\ref{eq6}) to (\ref{eq5}), as a result, the equation, which describes  spin evolution of the $\Omega^{\pm}$ baryon moving in a crystal, can be expressed as follows:
\begin{equation}\label{eq8}
\frac{d\hat{S}_i}{dt}=[\hat{\vec{S}}\times\vec{\Omega}_{magn}]_i+[\hat{\vec{S}}\times\vec{\Omega}_{EDM}]_i+
\frac{eQ}{3\hbar}\varepsilon_{ikl}\varphi_{kn}\hat{T}_{ln}.
\end{equation}

%
The third term in (\ref{eq8}), which is proportional to Q, includes second rank tensor $\hat{T}_{ln}$.
To find equations, describing time evolution of the tensor $\hat{T}_{ln}$ , let us use quantum mechanical Heisenberg equations \cite{LandauK}.

The operator equation describing the time evolution of an arbitrary quantum mechanical operator $\hat{A}$ can be written as follows:
\begin{equation}\label{eq9}
\frac{d\hat{A}}{d\tau}=\frac{i}{\hbar}\left[\hat{H},\hat{A}\right],
\end{equation}
where $\hat{H}$ is the Hamiltonian, $\left[\hat{H},\hat{A}\right]=\hat{H}\hat{A}-\hat{A}\hat{H}$ is the commutator.

As was stated above, in the high energy range the particle wavelength $\lambda$ is much smaller as compared to the typical field nonuniformity: $\lambda << R_{sh}$, where $R_{sh}$ is the atom shielding radius. Consequently, particle motion in crystals is quasi-classical, the particle is moving along the quasi-classical trajectory.

Let us consider interaction of a particle with electromagnetic field of the planes (axes) of the crystal in the particle's instantaneous rest frame. 
As a result of relativistic transformation, in this coordinate system electric field $\vec{E}_{\perp}$ increases and becomes $\vec{E}^{'}_{\perp} = \gamma \vec{E}_{\perp}$, appears magnetic field \\ $\vec{B}_{\perp}^{*'}=-\gamma[\vec{\beta}\times\vec{E}_{\perp}]=\gamma \vec{B}^{*}$, where $\gamma$ - Lorentz factor of the particle.
Energy of interaction between particles spin and magnetic field 
$ \hat{U}^{'}_{magn}=- \hat{\vec{\mu}}\vec{B}^{*'}_{\perp}$, \\ $ \mu = \frac{eg\hbar S}{2mc} \left( \frac{\hat{\vec{S}}}{S} \right) $.
Energy of interaction between  electric dipole moment and electric field  $U^{'}_{EDM}=-\hat{\vec{D}}\vec{E}^{'}$, $ \hat{\vec{D}} = ed\hbar\hat{\vec{S}}$.
Energy of interaction between quadrupole moment and field $ \vec{E}_{\perp} $ is 
$ \hat{U}^{'}_{a}=\gamma \frac{1}{6} eQ \varphi_{ln}\hat{T}_{ln}$.

At the first glance, the operator of interaction between particle and fields $\vec{E}^{'}$ and $\vec{B}^{*}$ denoted hereinbefore as $\hat{U}^{'}$  is the sum of $\hat{U}^{'}_{magn}$,   $\hat{U}^{'}_{EDM}$ and $\hat{U}^{'}_{Q}$. 
However, 
{when the particle  trajectory is not  linear}, the particle momentum rotates (cyclotron rotation).
This leads to occurrence of Thomas precession with frequency
 $\Omega_{T} = \frac{e}{mc}[\vec{\beta} \times E_{\perp}]$ \cite{Phys_13}. In the rest frame of the particle $\Omega^{'}_{T} = \gamma \Omega_{T}$. 
 
As a result of Thomas precession for a particle with magnetic moment that is equal to Bohr magneton, the angle between impulse and spin 
is not changing.
Term $\hat{U}^{'}_{T}$ 
{caused by Thomas precession} should be subtracted from $\hat{U}^{'}_{magn}$. This leads to replacement of $ g $ to $ g-2 $ in ${U}^{'}_{magn}$.
In this case energy of interaction between the spin and fields in the rest frame of the particle can be written as:
\begin{equation}\label{eq10}
\hat{U}^{'}= \gamma \Big\lbrace- \frac{e\hbar(g-2)}{2mc} \hat{\vec{S}} \vec{B}^{*}_{\perp} -ed\hat{S}\vec{E} + \frac{1}{6}eQ\varphi_{ln}\hat{T}_{ln}\Big\rbrace.
\end{equation}

In the rest frame of the particle equation (\ref{eq9}) is as follows:
\begin{equation}\label{eq11}
	\frac{d\hat{A}}{d\tau}=\frac{i}{\hbar}\left[\hat{U}^{'},\hat{A}\right].
\end{equation}

Let us replace $\tau$ with the time in laboratory frame $t = \gamma\tau$. As a result:
\begin{equation}\label{eq12}
	\frac{d\hat{A}}{dt}=\frac{i}{\hbar}\left[\hat{U},\hat{A}\right],
\end{equation}
where $ \hat{U}= \frac{\hat{U}^{'}}{\gamma} = -\frac{e \hbar (g-2)}{2mc} \vec{B}^{*}_{\perp}\hat{\vec{S}} -ed \vec{E}_{\perp} \hat{S} + \frac{1}{6} eQ\varphi_{ln}\hat{T}_{ln} $.

Lets consider equation (\ref{eq11}) for spin operator $ \hat{S}_{i} $:
\begin{equation}\label{eq13}
	\frac{d\hat{S}_{i}}{dt}=\frac{i}{\hbar}\left[\hat{U},\hat{S}_{i}\right],
\end{equation}
here i takes values 1,2,3 or x,y,z correspondingly in Cartesian coordinate system for spin operators $\hat{S}_{i}$. Using commutation relations for spin operators \cite{LandauK}:
\begin{equation}\label{eq14}
\left[\hat{S}_{l}, \hat{S}_{i}\right]=i \varepsilon_{lim}\hat{S}_{m},
\end{equation}
we can obtain:
\begin{equation}\label{eq15}
	\left[\vec{B}^{*}\hat{\vec{S}}, \hat{S}_{i}\right]=i \left[\vec{S} \times \vec{B}^{*}\right]_{i}, 
	\left[\vec{E}_{\perp}\hat{\vec{S}},\hat{S}_{i}\right]=i \left[\vec{S} \times \vec{E}_{\perp}\right]_{i}.
\end{equation}
\begin{equation}\label{eq16}
	\left[\hat{T}_{ln}, \hat{S}_{i}\right]=-i \varepsilon_{inm}\hat{T}_{lm}-i \varepsilon_{ilm}\hat{T}_{mn},
\end{equation}

\begin{equation}\label{eq17}
	\frac{1}{6} eQ \varphi_{ln}\left[\hat{T}_{ln},\hat{S}_{i}\right] = -i \frac{eQ}{3}\varepsilon_{ilm}\varphi_{ln}\hat{T}_{nm}.
\end{equation}
Let us insert (\ref{eq15}, \ref{eq17}) in equation (\ref{eq12}):
\begin{equation}\label{eq18}
\frac{d\hat{\vec{S}_{i}}}{dt}=\left[\vec{S} \times \vec{\Omega}\right]_{i} +
\frac{eQ}{3 \hbar} \varepsilon_{ilm} \varphi_{ln}\hat{T}_{nm},
\end{equation}
where $\vec{\Omega}= \vec{\Omega}_{magn} + \vec{\Omega}_{EDM}$.
If we omit the contribution caused by the quadrupole moment in equation (\ref{eq18}), we obtain equation (\ref{eq5}).

According to (\ref{eq18}) spin evolution in time depends not only on magnetic and electric dipole moments, but also on particle quadrupole moment Q.
Equation (\ref{eq12}) allows us to write equation, defining evolution in time of the components of quadropolarization tensor $\hat{T}$:
\begin{equation}\label{eq19}
\frac{d \hat{T}_{ik}}{dt} = \frac{i}{\hbar}[\hat{U},\hat{T}_{ik}].
\end{equation}
To obtain equation (\ref{eq19}) in its explicit form along with commutator (\ref{eq16}) commutator $ [\hat{T}_{ln}, \hat{T}_{ik}] $ is required.
With a bit bulky expressions we can obtain following equation for $ \varphi_{ln}[\hat{T}_{ln}, \hat{T}_{ik}] $:

\begin{eqnarray}\label{eq20}
&\varphi_{ln}[\hat{T}_{ln}, \hat{T}_{ik}]= i\varphi_{ln} [\varepsilon_{lkm}(S_i S_n + S_n S_i)S_m + \varepsilon_{lim} S_m (S_k S_n + S_n S_k) ] + \\ \nonumber
&+\varphi_{lk}S_i S_l -\varphi_{li}S_l S_k -\frac{1}{2} \varphi_{in} (S_k S_n + S_n S_k) + \frac{1}{2} \varphi_{kn} (S_i S_n + S_n S_i).
\end{eqnarray}
Commutators (\ref{eq16}) and (\ref{eq20}) allow us to obtain equations describing evolution of components $\hat{T}_{ik}$ of tensor $\hat{T}$:
\begin{eqnarray}\label{eq21}
\frac{d \hat{T}_{ik}}{dt} = 2 \varepsilon_{kml}\hat{T}_{mi} \Omega_l + 2 \varepsilon_{iml}\hat{T}_{mk} \Omega_l + \frac{i}{\hbar} \frac{eQ}{6} \varphi_{ln}[\hat{T}_{ln},\hat{T}_{ik}].
\end{eqnarray}

Commutator $ \varphi_{ln}[\hat{T}_{ln},\hat{T}_{ik}]$ included in (\ref{eq21}) is defined by equation (\ref{eq20}).
According to (\ref{eq21}) the derivative of $\hat{T}_{ik}$ depends not only on second rank tensor $\hat{T}_{ik}$, but also on the third rank tensor (see (\ref{eq20})).
We can obtain the explicit expression for the third rank tensor  using equation (\ref{eq12}), however it is not presented here due to bulkiness.

Let us consider planar channeling of the $\Omega^+$ baryon in crystal, bent around the x axis. Y axis is orthogonal to the family of planes in the  point of particle entrance into crystal. Z axis is orthogonal to the coordinate plane (x,y). Vector and tensor components on planes (x,y,z) correspond to the index values (1,2,3).
%
It is important to note that in instantaneous rest frame of the particle crystal is rotating around x axis, y axis is directed along the bending radius.
Over time particles spin is rotating regarding these axes.

In the laboratory frame the particle  spin rotates along the trajectory of particle motion in a crystal. 
After the particle enters the crystal with bending radius $R$, the electric field $\vec{E}_{\perp}$ between the crystal planes leads to movement of the channeled particle over the circular trajectory and rotation of the direction of particle momentum.
 Along with rotation of the particle impulse direction, rotation of the particle spin regarding impulse direction appears.
 Spin rotation in instantaneous rest frame of the particle is described by equations (\ref{eq18}) and (\ref{eq21}).
 In this frame, in each point of particle location the electric field $ \vec{E}_{\perp}$ is directed along the radius of crystal bending  and depends on bending radius value $R$.
Thus, symmetrical tensor $\varphi_{ln}$ in chosen coordinate system (x,y,z) has only one element with non-zero value $\varphi_{22} (\varphi_{yy})$.
Therefore we can write $\varphi_{ln}=\varphi_{22}\delta_{l2} \delta_{n2}$, where $\delta_{ln}$ is the Kronecker delta.
Now we can write equation (\ref{eq12}) as follows:

\begin{equation}\label{eq22}
	\frac{d\hat{S}_1}{dt}=-\Omega_{EDM}\hat{S}_3 + \frac{eQ}{3 \hbar} \varphi_{22} \hat{T}_{32}.
\end{equation}

\begin{equation}\label{eq23}
	\frac{d\hat{S}_2}{dt}=\Omega_{mag}\hat{S}_3.
\end{equation}

\begin{equation}\label{eq24}
	\frac{d\hat{S}_3}{dt}=-\Omega_{mag}\hat{S}_2 + \Omega_{EDM}\hat{S}_1 -\frac{eQ}{3 \hbar} \varphi_{22} \hat{T}_{12}.
\end{equation}
Beside components of the spin operator $ \hat{S}_i $, components of tensors $\hat{T}_{32}$ and $\hat{T}_{12}$, that affect $\hat{S}_i$ components changing in time, are also contained in equation (\ref{eq22}).
Equations (\ref{eq20},\ref{eq21}) lead to the following equation describing time dependence of operators $\hat{T}_{12}$ and $\hat{T}_{32}$:

\begin{equation}\label{eq25}
\frac{d \hat{T}_{12}}{dt}=2 \hat{T}_{31}\Omega_{magn}-2\hat{T}_{32}\Omega_{EDM} +\frac{eQ}{3 \hbar} \varphi_{22}\hat{S}_{2}\hat{T}_{23}+i\frac{eQ}{2 \hbar}\varphi_{22}\hat{T}_{12} - \frac{eQ}{6\hbar}\varphi_{22}\hat{S}_{3}.
\end{equation}

\begin{equation}\label{eq26}
	\frac{d \hat{T}_{32}}{dt}=2 \Omega_{magn}({T}_{33}-{T}_{22})+2\hat{T}_{12}\Omega_{EDM}-
	\frac{eQ}{3 \hbar}\varphi_{22}\hat{T}_{12}\hat{S}_{2}- i\frac{eQ}{2 \hbar}\varphi_{22}\hat{T}_{32} -\frac{eQ}{6\hbar}\varphi_{22}\hat{S}_{1}.
\end{equation}

According to (\ref{eq23}, \ref{eq24}), $\hat{T}_{12}$ and $\hat{T}_{32}$ changing in time depends on components of $\hat{\vec{S}_i}$. Besides this, $\hat{T}_{12}$ and $\hat{T}_{32}$ components depend on different components of the second rank tensor, and third rank tensors $\hat{S}_2\hat{T}_{32}$ and $\hat{T}_{12}\hat{S}_2 $. For them using equation (\ref{eq3}) we can obtain expressions describing their changing in time. However these equations are not quoted here due to bulkiness.
%
{After averaging the equations (\ref{eq22}-\ref{eq26}) with the spin density matrix, which
$\Omega^{\pm}$ baryon possesses before entering the crystal}, we obtain the similar expressions for polarization vector and polarization tensor of $\Omega^{\pm}$ baryon.
Lets use symbols without "hats" for averaged quantities  $S_i$, $T_{ik}$ ($S_i=Sp \hat{\rho}_s \hat{S}_i$, where $\hat{\rho}_s$ -spin density matrix).
Thus, as $\Omega^{\pm}$ baryon moves in crystal, not only spin rotation but also 
vector polarization transition into tensor polarization (and vice versa) appears due to quadrupole interaction.

Let us consider equation (\ref{eq22}) for components of spin vector $ \vec{S}_1 $. According to (\ref{eq22}) change in components of spin $\hat{S}_1$ appear only if EDM and Q are the non-zero values.
$\hat{S}_1$ changing in time depend on components of spin vector $ \hat{S}_3$ and components of tensor $\hat{T}_{32}$:
\begin{equation}\label{eq27}
\hat{S}_1(t)=\hat{S}_1(0) -\Omega_{EDM} \int_0^t \hat{S}_3(t')dt' + \frac{eQ}{3 \hbar}\varphi_{22}\int_0^t \hat{T}_{32}(t')dt',
\end{equation}
where $\hat{S}_1(0)$ is $\hat{S}_1$  value at the moment of particle entering the crystal.
If angle of spin rotation in crystal,caused by anomalous magnetic moment of the particle moving in crystal is much less than 1, than the approximate solution of (\ref{eq27}) is as follows:

\begin{equation}\label{eq28}
\hat{S}_1(t)=\hat{S}_1(0)
-\Omega_{EDM} \,t \,\hat{S}_3(0)+\frac{eQ}{3 \hbar}\varphi_{22} \,\, t\, \,\hat{T}_{32}(0),
\end{equation}
where $\hat{S}_1(0), \hat{S}_3(0), \hat{T}_{32}(0)$  are the spin characteristics
of the $\Omega^{\pm}$ at entering the crystal.

According to (\ref{eq27}), if component of the polarization vector directed along the x axis is equal to zero when particle enters the crystal, then appearance of such component means that either EDM or Q is a non-zero value.
To separate possible contributions to $\vec{S}_i(t)$ it is enough to compare measurements results for two crystals: one of which is bent clockwise, and another one that is bent counter-clockwise, or to compare measurements results for one crystal: in initial position and after turning the crystal 180° around the direction of impulse of the moving particle.

According to estimations \cite{Phys_1}, decrease of the $S_x$ ($S_1$) component  value due to presence of EDM  is $S_x \lesssim 10^{-2}$, when a heavy baryon passes 1 cm distance in bent Si crystal. 
Let us now estimate possible change in $S_1$ value due to presence of the qudrupole moment of $\Omega^{\pm}$ baryon.
According to (\ref{eq27}, \ref{eq28}) possible change in $S_1$ ($S_x$) value is determined by the parameter:
\begin{equation}\label{eq29}
\omega_{Q}= \frac{eQ\varphi_{22}}{3 \hbar}.
\end{equation}
%
Lets analyze the stated parameter when $\Omega^{+}$ is moving in non-bent crystal.
Poisson equation gives for $\varphi_{22}$:

\begin{equation}\label{eq:poisson}
\varphi_{22}=\frac{\partial^2 \varphi}{\partial y^2}= - 4 \pi \rho(y),
\end{equation}
where $\rho(y)=\rho_e(y)+\rho_n(y)$ is the density of charge in point $y$ averaged over the atoms distribution in the plane, $\rho_e(y)$ is the charge density for electrons,   $\rho_n(y)$ is the charge density for nuclei. Therefore, parameter $\omega$ reads as 
 \begin{equation}\label{eq:omega2}
 \omega=-\frac{4 \pi eQ}{3 \hbar} \rho(y).
 \end{equation}
When baryon moves far from nuclei positions (i.e. $\rho_n(y)=0$), the charge density $\rho(y)=\rho_e(y)=e n_e(y)$, where $n_e(y)$   is the density of electrons. Therefore, for a baryon moving between crystal planes this parameter is expressed as follows:
  \begin{equation}\label{eq:omega3}
 \omega=-\frac{4 \pi e^2Q}{3 \hbar} n_e(y),
 \end{equation}
i.e. 
 \begin{equation}\label{eq:omega4}
\omega= - \frac{4\pi}{3}\alpha Q c n_e(y),
\end{equation}
where $\alpha=\frac{e^2}{\hbar c}$ is the fine structure constant.

Let a particle move in the area, where interplane potential is minimal, The
 density of electrons can be evaluated as \cite{BOsmolovskiy}:  
\begin{equation}\label{eq:n_e}
n_e=\frac{n Z}{2.72},
\end{equation}
where $n$ is the density of atoms in the crystal, $Z$ is the nucleus charge.
Evaluation (\ref{eq:n_e}) for density of electrons in the vicinity of minimum of interplane potential corresponds to homogenious distribution of electrons with the density, which is $e$-times lower as compared to the average density of the crystal electrons $\overline{n}_e=Z n$.
Parameter $\varphi =|\omega| t$ defines change in the value of the polarization vector component $S_x$ (see (\ref{eq28})), here $t$ is the time during which particle moves in the crystal. 
Parameter $\varphi$ can be written as:
\begin{equation}\label{eq35}
\varphi = \frac{4 \pi}{3} \alpha Q \frac{nZ}{2.72}c \,t=\frac{4 \pi}{3} \alpha Q \frac{nZ}{2.72} L ,
\end{equation}
where L is the particle path length  in the crystal.
From (\ref{eq35}) we have:

\begin{equation}\label{eq36}
	\varphi \approx 10^{-2}QnZL.
\end{equation}
Evaluation of the $\varphi$ value for Si crystal ($Z=14$) of thickness $L= 10$ cm 
with the use of (\ref{eq36}) gives $\varphi_{Si}=1.4 10^{-5}$ (density of  crystal atoms is supposed to be $n=10^{22}, Q=10^{-27}$cm$^2$).
Similar evaluation for germanium (Z=32) gives $\varphi_{Ge}=3.2 10^{-5}$ at the same density of atoms. With the growth of $Z$ value, $\varphi$ value increases as well (\ref{eq36}).

The above evaluations are valid for a baryon moving at the minimum of the interplane potential.
When a baryon moves near the potential maximum the density $n_e(y)$ is several times higher (more than 3 times, see, for example calculations in \cite{Biryukov}).
%
Mind that by reducing the radius of curvature for a bent crystal one can move the equilibrium trajectory of positive baryons passing through the bent crystal to the area, where density $n_e(y)$ significantly exceeds value given by (\ref{eq21}) for particle traveling in the minimum of interplane potential \cite{Biryukov}.
Therefore, use of the bent crystal enables increasing of the effect.

It should be noted, that it also makes sense to conduct measurements of the $\Omega^{\pm}$ baryons quadrupole moment  in case of axial channeling and scattering by the axes  due to increase of the electric field, created by the axes. 

According to calculations made in \cite{Phys_1},
the increase of the electric field in case of axial channeling leads to the possibility of measuring MDM and EDM of heavy baryons.
Let us remind, that along with MDM, EDM and Q, study of  spin evolution for particles moving in a crystal opens possibility to study P-odd, T(CP)-odd interactions \cite{VG_6, VG_7,Bar_20}, 
as well as to observe the birefringence effect for $\Omega^{\pm}$ baryons (for the case of amorphous medium the birefringence effect was analyzed in \cite{Bar_Bar}).

    \section{Conclusion}
Experiments with $\Omega^{\pm}$ baryons moving in a bent and straight crystal at LHC in addition to EDM 
open possibility to study $\Omega^{\pm}$ baryons electric quadrupole moment, to observe the birefringers effect (spin oscillations and and dichroism) for $\Omega^{\pm}$ baryons and to get information about constants  describing P-odd, T(CP)-odd effects at $\Omega^{\pm}$ baryon interactions with electrons and nuclei.
Equations for spin and tensor polarization rotation are presented along with the estimations study for values of the effects under discussion.

\end{document}